\documentclass[10pt,journal,compsoc]{IEEEtran}
\ifCLASSOPTIONcompsoc
  \usepackage[nocompress]{cite}
\else
  \usepackage{cite}
\fi
\ifCLASSINFOpdf
\else
\fi
\usepackage[cmtip,all]{xy}
\usepackage[english]{babel}
\usepackage{amsmath,amssymb}
\usepackage{times}
\usepackage{relsize}
\usepackage{graphicx}
\usepackage{url}
\usepackage{booktabs}
\usepackage{multirow}
\usepackage{mparhack}
\usepackage{subfigure}
\usepackage{caption}
\usepackage{amsthm}
\usepackage{bm}
\usepackage{algorithm}
\usepackage[noend]{algorithmic}

\usepackage{array}
\usepackage{eqparbox}
\usepackage{mdwmath}

\usepackage{epsfig}
\usepackage{epstopdf}
\usepackage{xcolor}

\newtheorem{theorem}{Theorem}
\newtheorem{lemma}[theorem]{Lemma}

\newtheorem{corollary}[theorem]{Corollary}

\newcounter{exam}
\newcounter{rema}
\newtheorem{remark}[rema]{Remark}

\DeclareFontFamily{OT1}{pzc}{}
\DeclareFontShape{OT1}{pzc}{m}{it}%
              {<-> s * [1.1] pzcmi7t}{}
\DeclareMathAlphabet{\mathpzc}{OT1}{pzc}%
                                 {m}{it}

\def\CI{{\mathcal I}}

\def\CL{{\mathcal L}}
\def\CH{{\mathcal H}}

\def\CT{{\mathcal T}}

\usepackage{accents}
\newcommand\munderbar[1]{%
  \underaccent{\bar}{#1}}

\begin{document}
\captionsetup[figure]{name={Fig.},font=footnotesize,labelsep=period}
\title{Optimal Scheduling of Age-centric Caching: Tractability and Computation}

\author{Ghafour~Ahani,
        Di~Yuan,~\IEEEmembership{Senior Member,~IEEE,}
        and~Sumei~Sun,~\IEEEmembership{Fellow,~IEEE}
\IEEEcompsocitemizethanks{\IEEEcompsocthanksitem G. Ahani and D. Yuan are with the Department
of Information Technology, Uppsala University, 751 05 Uppsala, Sweden (e-mails: \{ghafour.ahani, di.yuan\}@it.uu.se).\protect\\
\IEEEcompsocthanksitem S. Sun is with the Institute for Infocomm Research, Singapore
(e-mail: sunsm@i2r.a-star.edu.sg).}
}
\IEEEtitleabstractindextext{%
\begin{abstract}
  The notion of age of information (AoI) has become an important
  performance metric in network and control systems. Information freshness, represented by AoI, naturally arises in the
  context of caching. We address optimal scheduling of cache updates
  for a time-slotted system where the contents vary in size.
  There is limited capacity for the cache and for making content
  updates.  Each content is associated with a utility function
  that is monotonically decreasing in the AoI.  For this combinatorial
  optimization problem, we present the following contributions. First,
  we provide theoretical results settling the boundary of
  problem tractability.  In particular, by a reformulation using
  network flows, we prove the boundary is essentially
  determined by whether or not the contents are of equal size.
  Second, we derive an integer linear formulation for the
  problem, of which the optimal solution can be obtained for
  small-scale scenarios. Next, via a mathematical reformulation, we
  derive a scalable optimization algorithm using repeated column
  generation.  In addition, the
  algorithm computes a bound of global optimum, that can be used to
  assess the performance of any scheduling solution. Performance
  evaluation of large-scale scenarios demonstrates the strengths of
  the algorithm in comparison to a greedy schedule. Finally, we extend
  the applicability of our work to cyclic scheduling.
\end{abstract}

}
\maketitle
\IEEEpeerreviewmaketitle
\section{Introduction}
The research interest in age of information (AoI) has been rapidly
growing in the recent years. The notion of AoI has been introduced for
characterizing the freshness of information~\cite{KaYaGr12}. AoI is
defined as the amount of time elapsed with respect to the time stamp
of the information received in the most recent update.
The AoI grows linearly between two successive updates. For a wide
range of control and network systems, e.g., status monitoring via
sensors, AoI has become an important performance metric.

The AoI aspect arises naturally in the context of caching for holding
content items with dynamic updates~\cite{YaCiYeWi17}. Consider a local
cache for which updates of content items take place via a network
such as a backhaul link in wireless systems. Rather
than obtaining content items via the backhaul network, users can download them
from the cache if the content items of interest are in the cache, thus
reducing network resource consumption. Due to limited backhaul network
capacity, not all content items in the cache can be updated
simultaneously. Hence, at a time point, the performance of caching
depends on not only the content items currently in the cache, but also
how old these items are; the latter can be characterized by AoI.

We address optimal scheduling of cache updates, where the utility of
the cache is based on AoI. The system under consideration is
time-slotted, with a given scheduling horizon.  The content items are
of different sizes. Updating the cache in a time slot is subject to a
capacity limit, constraining which content items that can be downloaded in the same time
slot. Moreover, the cache itself has a capacity. For each content
item, there is a utility function that is monotonically decreasing in
the AoI, and because the content items differ in popularity, the utility
function is content-specific. If a content item is not in the cache,
the utility value is zero. For every time slot, the optimization
decision consists of the selection of the content items to be updated,
subject to the network and cache capacity constraints. Thus the
problem falls in the domain of combinatorial optimization.  The
objective is to find the schedule maximizing the total utility over
the scheduling horizon.

Our work consists in the following contributions toward understanding
and solving the outlined AoI-driven cache optimization problem
(ACOP).

\begin{itemize}

\item We provide theoretical results of problem tractability.
Specifically, for the special case of
uniform content size, the global optimum of ACOP admits polynomial-time
tractability (even if the problem remains combinatorial optimization).
We establish this result via mapping the problem to a graph, and proving
that the optimal schedule is a minimum-cost flow in the graph.  For
non-uniform content size, the problem is NP-hard due to the knapsack
structure, except for two contents. Thus our results settle the boundary
of problem tractability.

\item We derive an integer linear programming (ILP) formulation
for ACOP in its general form, enabling the use of off-the-shelf
optimization solvers to approach the problem.  This is particularly
useful for examining the performance of sub-optimal solutions for
small-scale scenarios, for which the global
optimum is computed via ILP.

\item  We derive a mathematical reformulation,
by considering sequences representing the evolution of the AoI
over time for each content item. The reformulation enables a scalable solution
approach. Specifically, we present a column generation algorithm that
addresses the linear programming (LP) version of the reformulation, by
keeping only a small subset of possible sequences and augmenting the
subset based on optimality condition. To obtain integrality,
we present a rounding concept based on disjunctions rather
than on fractional variables, such that column generation is
repeated for improvement after rounding.  Performance
evaluation demonstrates the strengths of the repeated column
generation algorithm in comparison to a greedy schedule.  Moreover, as
the algorithm provides an optimality bound in addition to a problem
solution, it is possible to gauge performance even if
the global optimum is out of reach. Using the bound, our results show the
algorithm consistently yields near-to-optimal solutions for
large-scale problem instances.

\item Finally, we discuss the extension of our results
to cyclic scheduling, i.e., the schedule is repeated in a cyclic
manner.  In this case, the AoI values at the beginning of the schedule
are determined by the caching updates made later in the schedule.  We
present adaptations, such that the
tractability analysis, the ILP formulation, and the repeated
column generation algorithm all remain applicable.

\end{itemize}

\section{Related Work}
\label{sec:related}

The notion of AoI was introduced in \cite{KaYaGr12}.
Generalizations to multiple information sources have been studied in
\cite{YaKa12}.  The aspect of queue management has been
introduced in \cite{KakoEp13,CoCoEp14,PaGuKrKoAn15}.  Early
application scenarios of AoI include channel
information~\cite{CoVaEp14} and energy harvesting~\cite{Ya15}. In
\cite{SuBiYaKoSh17}, the authors address optimal update policy for
AoI, and provide conditions under which the so called zero-wait policy
is optimal. In \cite{SaLiJi17}, the authors have considered AoI under a
pull model, where information freshness is relevant (only) when the
use interest arises.  For transmission scheduling with AoI
minimization, \cite{KaUySiMo16} considers a system model with error
probability of transmission, and \cite{HeYuEp18} proposes algorithms
for multiple link scheduling with presence of interference. We refer
to \cite{KoPaAn17} for a comprehensive survey of research on AoI.
Below we outline very recent developments that demonstrate a rapidly
growing interest in the topic.

One line of research has consisted in sampling, scheduling and
updating policies for various system models that have
queuing components and AoI as the performance objective.
In \cite{SuBiKo18}, the scenario has multiple
source-destination pairs with a common queue, along with a scheduler.
In \cite{KaKoNgWiEp18}, the authors have investigated optimal sampling
strategies addressing AoI as well as error in estimating the state of
a Markov source.  The system model in \cite{SuJiZhNi19} combines
stochastic status updates and unreliable links; the authors make an
approximation of the Whittle index and derive a solution algorithm
thereby. The study in \cite{FaKlBr18} considers an energy-constrained
server that is able to harvest energy when no packet from the source
requires service. The notion of preemption has been addressed in
\cite{Ya18,NaTe18} for a single flow in a multi-hop network
with preemptive intermediate nodes, and multiple flows such that
different flows preempt each other, respectively.  The authors of
\cite{WaDu19} have proposed the use of dynamic pricing to provide
AoI-dependent incentives to sampling and transmission.
For network control systems (NCPs),
\cite{ChMaJoGr19} has addressed sampling and scheduling, relating estimation error to AoI,
and \cite{KlMaHiKe19} has focused on approximately optimal scheduling
strategy for multi-loop NCPs, where the centralized scheduler takes a
decision based on observed AoI.

Another line of research arises from the introduction of AoI to
various applications and networking context. By including
energy constraints, \cite{HeDaFo19} extends \cite{HeYuEp18} for
optimal link activation for AoI minimization in wireless networks
with interference.  The study in \cite{FaKlBr19} is similar to
\cite{HeYuEp18} and \cite{HeDaFo19} in terms of the scheduling aspect and the presence of
interference, however the system model considers AoI for all node pairs of the network, and the
emphasis is on performance bounds. The work in
\cite{DoChFa18} considers a two-way data exchanging
system, where the AoI is constrained by the uplink transmission capability and the downlink energy transfer capability.
Application of the AoI concept in camera networks where information
from multiple cameras are correlated has been presented in
\cite{HeDaFo18}. For remote sensor scenarios,
\cite{LiWaBaDa18} has studied AoI-optimal trajectory planning for
unmanned aerial vehicles (UAVs).
For distributed content storage, AoI has been applied
to address the trade-off between delay and data staleness
\cite{ZhYaSo18}. In \cite{MaAsEp19}, the authors have studied
AoI with respect to orthogonal multiple access (OMA) and
non-orthogonal multiple access (NOMA), and revealed that
NOMA, even though allowing for better spectral efficiency,
is not always better than OMA in terms of average AoI.

AoI has also appeared as the performance metric in using machine learning
as a tool in communication systems. In \cite{CeGuGy18}, online adaptive learning
has been used for addressing error probability in the context of AoI
minimization. In \cite{KaKoEp19}, learning is used for
optimal data sampling to minimize AoI.

For AoI-aware caching, research results are available in
\cite{YaCiYeWi17,TaCiWaWiYa19}. The general problem setup, i.e.,
what to optimize and when to update with an
objective function defined by AoI,
reminds the system model we consider. However, our work has
significant differences to \cite{YaCiYeWi17} and \cite{TaCiWaWiYa19}.
First, it is (implicitly) assumed
in \cite{YaCiYeWi17} and \cite{TaCiWaWiYa19} that the items are of uniform
size.  We do not have this
restriction and hence in our problem,
which items can be updated at a time are determined by the
sizes as well as the capacity limits.
Second, the works in
\cite{YaCiYeWi17} and \cite{TaCiWaWiYa19} derive updating policy with respect to
expected AoI, whereas our problem falls into the domain of
combinatorial optimization and we use methods thereof to solve the
problem. Moreover, the problems in \cite{YaCiYeWi17,TaCiWaWiYa19} are
approached by either assuming given inter-update intervals of each item or
the total number of updates of each item. In our work, these entities
remain optimization variables throughout the optimization process.

\section{Preliminaries}

\subsection{System Model}
\label{sec:system}

Consider a cache of capacity $C$. The content items for caching
form a set $\CI = \{1, \dots, I\}$. Item $i \in \CI$ is of size $s_i$.
Time is slotted, and the time horizon consists of a
set of time slots $\CT = \{1, \dots T\}$. In each time slot, the cache
can be updated via a backhaul communication link whose capacity is
denoted by $L$.

The AoI of an item in the cache is the time difference between the
current slot and the time slot in which the item was most recently updated.  Each
time the item is updated, the AoI is zero, i.e., maximum information
freshness. The AoI then increases by one for each time slot, until the
item gets updated again.  The value of having item $i$ cached is
characterized by a utility function $f_i(\cdot)$ that is monotonically
decreasing in its AoI. If the item is not in the cache,
its utility is zero.  The utility function is item-specific to
reflect, for example, the difference in the popularity of the content items.

The AoI-driven cache optimization problem, or ACOP in short, is to
determine which content items to store and update in each time slot, such that
the total utility of the cache over the time horizon $1 \dots
T$ is maximized, subject to the capacity limits of the cache and
the backhaul. Later in Section~\ref{sec:cyclic}, we will consider
the case of optimizing a cyclic schedule of cache updates.

Notation: In addition to regular mathematical style of entities, we adopt the
following notation style in the paper. Sets are denoted using
calligraphic style. Moreover, boldface is used to denote
the vector form of the entity in question.

\begin{remark}
Our system model is not necessarily restricted to caching scenarios.
For example, consider a system monitoring a number of remote sites,
for which the information generated differ in size.  The amount of
information that can be sent to the monitoring center is constrained by the
network bandwidth, and the task is optimal
scheduling of updates to maximize utility as a function of AoI. This setup
corresponds to ACOP with redundant cache capacity. $\square$
\end{remark}

\subsection{Greedy Solution}
\label{sec:greedy}

For a combinatorial problem, a simple and greedy strategy typically
serves as a reference solution. For ACOP, a greedy solution is to
maximize the total utility of each time slot by simply considering the
utilities of the individual items if they are added to the cache or
become updated in the cache. This solution for a generic time slot is given in
Algorithm~\ref{alg:greedy}.

\begin{algorithm}[ht!]
{\bf Input}: $\CI$, $s_i, i \in \CI$, current cache $\CH$, current AoI $a_i, i \in \CI$ \\
{\bf Output}: $\CH$ \\
\vspace{-4mm}
\caption{Greedy solution for a generic time slot}
\label{alg:greedy}
\begin{algorithmic}[1]
\STATE $\CI' \gets \CI$; $L' \gets L$; $r \gets C - \sum_{i \in \CH} s_i$; $\CH' \gets \emptyset$
\WHILE {$\CI' \not= \emptyset$ and $L' > 0$}
\STATE $i^* \gets \text{argmax}_{i \in \CI'} f_i(0)$
\STATE $\CI' \gets \CI' \setminus \{i^*\}$
  \IF {$i^* \in \CH$}
      \STATE Update cache item $i^* \in \CH$
      \STATE $\CH' \gets \CH' \cup {i^*}$
  \ELSE
      \IF {$r + \sum_{i \in \CH \setminus \CH'} s_i \geq s_{i^*}$ and $s_{i^*} \leq L'$}
         \WHILE {$r<s_{i^*}$}
            \STATE $i' \gets \text{argmin}_{i \in \CH} f_i(a_i) $
            \STATE $\CH \gets \CH \setminus {i'}$
            \STATE $r \gets r + s_{i'}$
         \ENDWHILE
         \STATE $\CH \gets \CH \cup \{i^*\}$
         \STATE $r \gets r - s_{i^*}$
         \STATE $L' \gets L' - s_{i^*}$
      \ENDIF
  \ENDIF
\ENDWHILE
\end{algorithmic}
\end{algorithm}

In the algorithm, $\CI'$ is the candidate set of items for caching,
$r$ is the residual cache capacity, and $\CH'$ is used to keep track of
those items that are currently in the cache and selected for
update. The item $i^*$ maximizing the utility, if added or updated, is
considered and then removed from $\CI'$. If $i^*$ is in the cache, it
gets updated and recorded in $\CH'$.  Otherwise, if the remaining downloading
capacity admits and there will be sufficient residual capacity by
removing cached items except those in $\CH'$, the algorithm
removes items in ascending order of their utility values with respect
to the current AoI, followed by adding item $i^*$. The process ends
when the candidate set $\CI'$ becomes empty.

\section{Problem Complexity Analysis}
\label{sec:complexity}

ACOP apparently is in the domain of combinatorial optimization.
Thus it is important to gain understanding of problem complexity, as many
combinatorial problems are NP-hard (e.g., the traveling salesman
problem) whereas others are tractable in terms of computing global
optimum in polynomial time (e.g., matching in graphs).
In this section, we present proofs to establish the boundary of
tractability for ACOP. Namely, for uniform item size, ACOP is
tractable, otherwise it is NP-hard except for the very special case of
two content items. We denote ACOP with uniform
item size by ACOP$_\text{u}$.

As the capacity limits of the cache and the backhaul link
imply constraints of knapsack type, it is not surprising
that ACOP is NP-hard in general, with a proof based on the binary
knapsack problem. This result is formalized below.

\begin{theorem}
\label{theo:nphard}
ACOP is NP-hard.
\end{theorem}
\begin{IEEEproof}
Consider a knapsack problem with $N$ items and knapsack capacity $B$.
Denote the value and weight of item $i$ by $v_i$ and $w_i$,
respectively. We construct an ACOP instance by setting $T = 1$ (i.e.,
single time slot), $I = N$, $L = C = B$, and $s_i = w_i, i=1,\dots,I$.
The utility function is defined such that $f_i(0) = v_i, i=1,\dots,I$.
Downloading a content item to the cache amounts to selecting the
corresponding item in the knapsack problem. Obviously, the optimal
solution maximizing the total utility of the cache leads to
the optimum of the knapsack problem, and the result follows.
\end{IEEEproof}

Consider ACOP$_\text{u}$ where all items are of the same size. In this
case, both cache and backhaul capacities can be expressed in the
number of items. Even though the problem is still combinatorial along the time
dimension, we prove it is tractable, i.e., the global optimum can be
computed in polynomial time. The key is to transform the problem into
a minimum-cost flow problem ~\cite{netflow} in a specifically
constructed graph. In the following, we assume that the backhaul capacity
$L$ is non-redundant, i.e., $L < C$.

In optimization, a network flow problem is defined in a (directed) graph; each arc has
a linear cost in the arc flow that is also subject to an upper bound
and a lower bound. The latter is often zero. There are one or more
source nodes and sink nodes, each generating and receiving a specified
amount of flow, respectively. The total amount of flow generated at
the source nodes equals that to be received by the sink nodes.  The
optimization task is to determine how flows go from the source nodes
to the sink nodes, such that the total flow cost is minimum, subject
to flow balance at the nodes as well as the upper and lower bounds of
the arcs.

\begin{figure*}
\centering
  \includegraphics[width=0.7\textwidth,height=7.2cm]{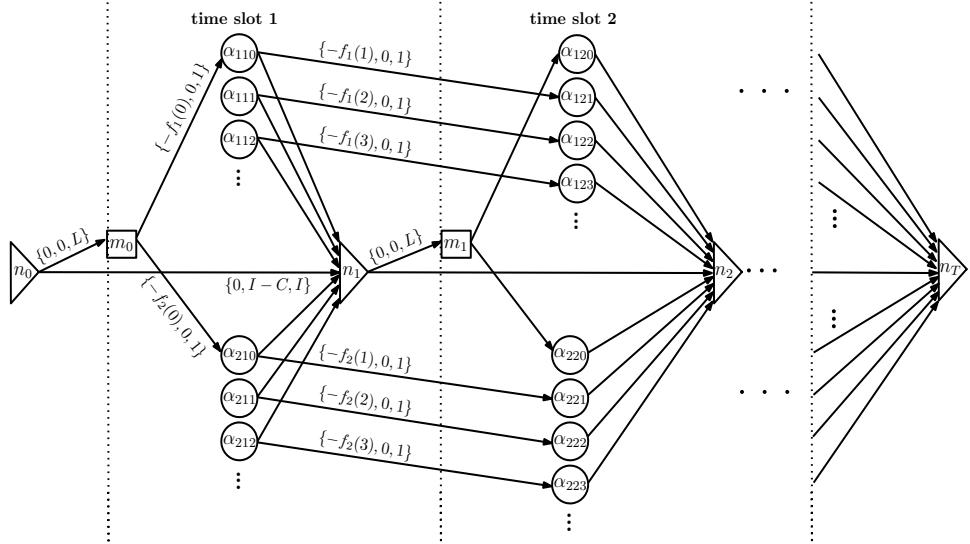}
  \caption{An illustration of the network flow problem corresponding to ACOP$_\text{u}$.}
  \label{fig:flow}
\end{figure*}

The graph we construct for proving the tractability of ACOP$_\text{u}$
is illustrated in Fig.~\ref{fig:flow}. For clarity, we illustrate
the construction for two items and the first two time slots.  The
construction of the remaining time slots follows the same pattern.
There is one source node, $n_0$,
and one sink node $n_T$. The source node generates $I$ flow units to
be sent to the sink through the network.

The underlying rationale is as follows.  There are
three types of nodes. First, for each item $i$ and time slot
$t$, node $\alpha_{ita}$ represents that item $i$ is in the cache in slot
$t$ with age $a$. Second, nodes $n_t, t=1, \dots, T$ act as the collection
points for all items except those to be kept in the cache in time slot
$t$.  Third, nodes $m_t, t=1, \dots, T-1$, are collection points for items to
be added or updated in the cache.

Each arc of the graph is associated with tuple $(c, l, u)$, where $c$
is the cost per flow unit, and $l$ and $u$ are the lower and upper
bounds of the arc flow, respectively.  These values are shown in the
figure, however not for the arcs entering the nodes $n_t, t=1, \dots,
T$ due to lack of space.  For these arcs, the tuple is $(0,0,1)$.
Moreover, we do not show explicitly the tuples for some arcs of time
slot two for the sake of clarify; the tuple values of any such arc are
the same as for the corresponding arc in the previous time slot.

Let us consider the first time slot with any integer flow solution.
The $I$ flow units leaving source $n_0$ have to either enter $m_0$ or
$n_1$.  Note the maximum number of units entering $m_0$ has upper
bound $L$, allowing at most $L$ items to be added or updated.  Node
$m_0$ has an arc of capacity one to node $\alpha_{i10}$ of item $i$,
$i=1,\dots,I$.  Having a flow of one unit on the arc corresponds to
putting item $i$ in the cache, generating utility $f_i(0)$ that is
equivalent to an amount of $-f_i(0)$ in cost minimization. There will
be $L$ such items\footnote{For ACOP$_\text{u}$, the downloading capacity $L$ is always
fully used, because adding or updating items always leads to better
utility (or more negative cost for the minimum-cost flow problem).}. The
flow units from $n_0$ to $n_1$ correspond to the items not cached in
time slot one. Note the flow lower bound of arc $(n_0, n_1)$ is $I-C$,
meaning that at least $I-C$ flow units must enter node $n_1$, i.e.,
at least $I-C$ items are outside the cache.

For a generic time slot $t$ and item $i$, if one unit of flow
enters node $\alpha_{ita}$, then the flow has two choices by graph
construction. Either it goes to node $\alpha_{i(t+1)(a+1)}$,
representing that item $i$ remains in the cache, with AoI $a+1$
for the next time slot and the corresponding utility, or it has to be
sent to the collection node $n_{t+1}$.  The latter case represents
that the item is removed from the cache, and, from $n_{t+1}$, the
item either enters $\alpha_{i(t+1)0}$ via node $m_{t+1}$ (i.e., item $i$ is
downloaded again to the cache with AoI zero), or stays outside the
cache by entering collection node $n_2$.

\begin{remark}
In  Fig.~\ref{fig:flow}, some of the nodes (and hence also their adjacent
arcs) are redundant. For example, no flow will enter nodes $\alpha_{122}$
and $\alpha_{123}$, because the AoI will never attain two or three in time slot
two. These nodes are however kept in the figure to better reflect the general structure of
graph construction. $\square$
\end{remark}

\begin{lemma}
\label{theo:toflow}
The optimal solution to ACOP$_\text{u}$ corresponds to an integer flow
solution for the constructed graph with equivalent objective function
value.
\end{lemma}
\begin{IEEEproof}
Consider an optimal solution to ACOP$_\text{u}$. For time slot one,
$L$ items are downloaded, and the flow on arc $(n_0, m_0)$ is set to
$L$, whereas $I-L$ units of flow are put on $(n_0, n_1)$.  For each
downloaded item $i$, we set the flow on $(m_0, \alpha_{i10})$ to be exactly
one unit, generating the (negative) cost of $-f_i(0)$.

Consider a generic time slot $t$. For the next time slot $t+1$, denote
the number of deleted items by $I_d$, and the numbers of items updated
and added by $I_u$ and $I_e$, respectively. Denote by ${\hat I}$ the
number of items not in the cache in the ACOP$_\text{u}$ solution for
time slot $t$.  Thus there are $\hat I$ flow units on arc $(n_{t-1},
n_t)$. If a cached item $i$ in the ACOP$_\text{u}$ solution is kept in
the cache in $t+1$, we set one flow unit on the arc representing this
state.  That is, if the AoI of $i$ is $a$, we set one flow unit on arc
$(\alpha_{ita}, \alpha_{i(t+1)(a+1)})$, generating a (negative) cost of
$-f_i(a+1)$. For any item $i$ that is either to be deleted or updated
for $t+1$, we set one flow unit on arc $(\alpha_{ita}, n_{t})$.  Thus
there are $I_d+I_u$ flow units in total arriving $n_t$ via these arcs,
and the total amount of incoming flow to $n_t$ equals $I_d + I_u +
{\hat I}$.

We set $L$ flow units on arc $(n_t, m_t)$. Next, observe that $L = I_e
+ I_u$, because at optimum the backhaul capacity is always fully
utilized. From $m_t$, for any item $i$ that is either updated or added
to the cache in $t+1$, we set one unit flow on arc $(m_t, \alpha_{i(t+1)0})$,
giving the (negative) cost of $-f_i(0)$.  By flow balance, the amount of
outgoing flow on arc $(n_{t}, n_{t+1})$ equals $I_n = {\hat I} + I_d +
I_u - L ={\hat I} + I_d + I_u - I_e - I_u = {\hat I} + I_d - I_e$, and
we need to prove $I_n$ is between the flow bounds $I-C$ and $I$.  If the
cache is full, then clearly $I_e=I_d$, and the conclusion follows.
Suppose the cache is not full (this occurs at the first few time
slots if downloading capacity $L$ is much smaller than cache capacity
$C$), such that the spare capacity can hold $\Delta$ items.  In
this case, at optimum, if $I_d > 0$, then each deleted item
will be replaced by an added item, as otherwise it is optimal
not to delete.  Let $\Delta' = \Delta + I_d$. We have
$I_e = \min \{L, \Delta' \}$.
Moreover, as $C- \Delta$ items are cached, the number of items outside
the cache ${\hat I} = I - C + \Delta$.  Thus $I_n = {\hat I}+I_d - I_e
= I - C + \Delta + I_d - \min\{L, \Delta'\}$ = $I - C + \Delta' - \min\{L, \Delta'\}$.
Note that $\Delta' - \min\{L,
\Delta'\} \geq 0$, thus $I_n \geq I-C$.  That $I_n \leq I$ follows
simply from that $\Delta' \leq C$.

In addition to the above, it is straightforward to see that the flow
construction satisfies the flow balance at nodes representing the
items' possible AoI values, and adheres to the bounds of their
adjacent arcs. At the last stage, for all cached items in slot $T$, we
set one unit of flow from the node representing the AoI to $n_T$.
These flow units, together with those on arc $(n_{T-1}, n_T)$ that represent
the number of items outside the cache, arrive the destination node
$n_T$ with a total flow of $I$. Hence the lemma.
\end{IEEEproof}

\begin{lemma}
\label{theo:fromflow}
The optimal integer flow solution in the constructed graph
corresponds to a solution of ACOP$_\text{u}$ with
equivalent objective function value.
\end{lemma}
\begin{IEEEproof}
For any integer flow solution, all arcs adjacent to nodes
$a_{it\alpha}, i =1, \dots, I, t=1,\dots, T,  a = 1,\dots, T-1$,
have either zero or one unit of flow. Moreover, obviously an optimal
flow will have $L$ flow units on arcs $(n_{t-1}, m_{t-1}), t=1,\dots,
T$.

Consider the flows on the outgoing arcs of node $m_0$.  Exactly $L$
arcs have one unit of flow, and the other arcs have zero flow.
For any arc $(m_0, \alpha_{i10})$ with one flow unit, the corresponding
solution of ACOP$_\text{u}$ downloads and caches item $i$ for time
slot one. Doing so for all time slots gives the ACOP$_\text{u}$ solution in
terms of the items that are added to or updated in the cache over time.
Moreover, any arc $(\alpha_{ita}, \alpha_{i(t+1)(a+1)})$ with a flow unit
means to keep the item $i$ in the cache from $t$ to $t+1$.  Thus the
flow solution leads to a caching solution for ACOP$_\text{u}$.  Note
that the solution is feasible with respect to cache capacity.  This is
because there are at least $I-C$ flow units on arc $(n_{t-1}, n_t)$,
and by flow balance, for any time slot $t, t=1,\dots, T$, there
are at most $C$ flow units in total on the incoming arcs of
$\alpha_{ita}, i =1, \dots, I, t=1,\dots, T,  a = 1,\dots, T-1$.
Finally, the constructed solution clearly gives an objective function
value that equals the negation of the optimal flow cost, and the lemma follows.
\end{IEEEproof}

\begin{theorem}
\label{theo:uniform}
ACOP$_\text{u}$ is tractable with polynomial-time complexity.
\end{theorem}
\begin{IEEEproof}
The maximum possible age of any item in the cache is bounded by the
number of time slots $T$. Hence the size of the constructed graph is
polynomial in $I$ and $T$.  Moreover, the graph is clearly acyclic.
For minimum-cost flow problems with integer input, there is an optimal
solution in which the arc flows are integers, and there are (strong)
polynomial algorithm for the problem including that
with negative costs in an acyclic graph\cite{netflow}.  These facts,
together with Lemmas~\ref{theo:toflow}-\ref{theo:fromflow}, establish
the theorem.
\end{IEEEproof}

Consider now ACOP with two content items. For this special case, it is
trivial to see if the cache can hold one or both items.  Moreover,
without loss of generality, one can assume that the backhaul capacity
allows for updating one item. As a result, ACOP with two content items
falls in the domain of ACOP$_\text{u}$, giving the corollary below.

\begin{corollary}
ACOP with two items is tractable with polynomial-time complexity.
\end{corollary}

The above observation does not generalize to three or
more items, because then the capacity can no longer be interpreted in
the number of items. Instead, the items' individual sizes
must be explicitly accounted for.

\begin{remark}
For combinatorial optimization problems, tractable problem
sub-classes are often identified by proving that a greedy solution leads
to optimum.  For ACOP$_\text{u}$, however, the greedy solution in
Section~\ref{sec:system} remains sub-optimal. This is due to
  item-specific utility function and backhaul capacity. Consider
  a simple example of two items.  Both are of size one. The
  capacity values are $C=2$ and $L=1$.  For item one, the utility
  function $f_1(0) = 2k$ (with $k>1$) and $f_1(a) = 0$ for $a \geq 1$. For
  item two, $f_2(a) = k$ for $a \leq t$, and $f_2(a) = 0$ for $a \geq t+1$.
  The greedy solution would cache and update item one in all time
  slots. Thus for $T = t$, the total utility is $2kt$. The
  optimal solution is to cache item two in time slot one.  Then,
  item one is added and kept updated in the next $t-1$ time slots.
  This gives a total utility of
  $kt + 2k(t-1) = 3kt-2k$. As a result, the greedy solution becomes
  highly sub-optimal for large $t$. Thus the notion of network flows
  is necessary for arriving at the conclusion of the tractability of
  ACOP$_\text{u}$. $\square$
\end{remark}

\section{Integer Linear Programming Formulation}
\label{sec:integer}

We derive an integer linear programming (ILP) formulation for ACOP.
This leads to a solution approach of using an off-the-shelf
optimization solver (e.g.,\cite{Gurobi-2019}).  Even though the
approach does not scale, it can be used to obtain optimum problem
instances of small size, for the purpose of performance evaluation of
other algorithms.

We use binary variable $x_{ita}$ that equals one if the age of item
$i$ in time slot $t$ is $a$, otherwise the variable equals zero. Note
that index $a$ is up to $t-1$, because the AoI in slot $t$ can never
exceed $t-1$. Also, $x_{it0}=1$ means item $i$ is added to the cache or
updated for time slot $t$. Binary variable $y_{it}$ is used to
indicate if item $i$ is in the cache in time slot $t$ or not. The ILP
formulation is given below.

\begin{subequations}
\label{eq:ilp}
\begin{align}
\label{eq:ilpobj}
  \qquad & \hspace{-4mm} \max_{\bm{x}, \bm{y}} \limits\
   \sum_{i \in \CI}\sum_{t \in \CT} \sum_{a=0}^{t-1} f_i(a) x_{ita} \qquad &  \\
\label{eq:ilpstate}
\qquad &   \text{s.t.}  \ \
\sum_{a=0}^{t-1} x_{ita} = y_{it},  i \in \CI,\ t \in \CT  &\\
\label{eq:ilpnextslot}
\qquad &
\hspace{2mm} (1-y_{i(t+1)}) + x_{i(t+1)0} + x_{i(t+1)(a+1)} \geq x_{ita},  \ \nonumber \\
& ~~~~~~~~~~ i \in \CI,t=1,\dots,T-1, a=1,\dots,t-1 &\\
\label{eq:ilpccap}
\qquad &
\hspace{2mm} \sum_{i \in \CI} s_i y_{it} \leq C,  \ t \in \CT & \\
\label{eq:ilpbcap}
\qquad &
\hspace{2mm} \sum_{i \in \CI} s_i x_{it0} \leq L,  \ t \in \CT & \\
\label{eq:ilpxdomain}
\qquad &
\hspace{2mm} x_{ita} \in \{0, 1\},  i \in \CI, \ t \in \CT,a=0,\dots,t-1 & \\
\label{eq:ilpydomain}
\qquad &
\hspace{2mm} y_{it} \in \{0, 1\},  i \in \CI,\ t \in \CT &
\end{align}
\end{subequations}

The objective function \eqref{eq:ilpobj} is to maximize the overall
utility.  By \eqref{eq:ilpstate}, if item $i$ is cached in a time slot
$t$, i.e., $y_{it}=1$, then exactly one of the binary variables
representing the possible AoI values is one, otherwise these binary
variables have to be zeros.  Inequality \eqref{eq:ilpnextslot}
states that if item $i$ is cached with AoI $a$ in a time slot, then for
the next time slot, either the AoI becomes $a+1$ (represented by
$x_{i(t+1)(a+1)}=1$), or it gets updated (represented by
$x_{i(t+1)0}=1$), or the item is no longer in the cache (represented by
$y_{it}=0$). The cache and backhaul capacity limits are formulated in
\eqref{eq:ilpccap} and \eqref{eq:ilpbcap}, respectively.

\section{Repeated Column Generation Algorithm}
\label{sec:column}

For efficient solution of ACOP, we propose an
algorithm based on repeated column generation. Column generation is an
efficient method for solving large-scale linear programs with the following two
structural properties \cite{LUDe04}.  First, there are exponentially
many columns (i.e., variables), hence including all is not practically
feasible and the method deals with only a small subset of them.
Second, identifying new columns that improve the objective
function value can be performed by solving an auxiliary problem, named the
subproblem.  This enables successive addition of new and promising
columns until optimality is reached.

\subsection{Problem Reformulation}

Applying column generation to ACOP is based on a reformulation. In the
reformulation, a column of any item is a vector representing the
caching and updating decisions of the item over all time slots.  We
denote by $\CL_i$ the index set of all possible such vectors for item $i$.  For
each column $\ell \in \CL_i$, there is a tuple $(b_{it\ell}, u_{it\ell})$
for every time slot $t$. Both elements are binary.  Specifically,
a column is represented by $[(b_{i1\ell}, u_{i1\ell}), (b_{i2\ell}, u_{i2\ell}), \dots,
(b_{iT\ell}, u_{iT\ell})]$ in which
$b_{it\ell}$ and $u_{it\ell}$ are one if and only if the item $i$ is cached
and updated, respectively, in time slot $t$.  Note that there are
exactly three possible outcomes of the tuple values: $(0,0)$, $(1,0)$,
and $(1,1)$. As a result, the cardinality of set $\CL_i$ is
bounded by $3^T$.

Because a column $\ell$ fully specifies the caching and updating
solution, the associated AoI values over time are known for any given
column.  Hence,  the total utility
value of column $\ell$ for item $i$, denoted by $f_{i\ell}$, can be computed.

The problem reformulation uses a binary variable $\chi_{i\ell}$
for each item $i \in \CI$ and column $\ell \in \CL$.  This variable is
one if column $\ell$ is selected for item $i$, and zero otherwise.
The reformulation is given below.

\begin{subequations}
\label{eq:reform}
\begin{align}
\label{eq:reformobj}
  \qquad & \hspace{-4mm} \max_{\bm{\chi}} \limits\
  \sum_{i \in \CI} \sum_{\ell \in \CL} f_{i\ell} \chi_{i\ell} \qquad &  \\
\label{eq:reformone}
\qquad &   \text{s.t.}  \ \
\sum_{\ell \in \CL} \chi_{i\ell} = 1, \ i \in \CI  & \\
\label{eq:reformccap}
\qquad &
\hspace{2mm} \sum_{i \in \CI} \sum_{\ell \in \CL} b_{it\ell} s_i \chi_{i\ell} \leq C, \ t \in \CT  & \\
\label{eq:reformbcap}
\qquad &
\hspace{2mm} \sum_{i \in \CI} \sum_{\ell \in \CL} u_{it\ell} s_i \chi_{i\ell} \leq L, \ t \in \CT  & \\
\label{eq:reformchidomain}
\qquad &
\hspace{2mm} \chi_{i\ell} \in \{0,1\}, \ i \in \CI, \ell \in \CL_i  &
\end{align}
\end{subequations}

In \eqref{eq:reform}, \eqref{eq:reformone} states that exactly one
column (i.e., caching and updating solution) has to be selected for
every item.  The next two constraints formulate the cache and backhaul
capacity, respectively. Note that both
\eqref{eq:reform} and \eqref{eq:ilp}
are valid optimization formulations of ACOP. However they differ in
structure.

\subsection{Column Generation}
\label{section:column}

The structure of \eqref{eq:reform} can be exploited by using
column generation. To this end, we consider the linear
programming counterpart of \eqref{eq:reform}, where
\eqref{eq:reformchidomain} is replaced by the relaxation
$0 \leq \chi_{i\ell} \leq 1, i \in \CI, \ell \in \CL$.  Moreover, for
each item $i$, a small subset $\CL_i^\prime \subset \CL_i$ is used and
successively augmented to approach optimality.  Initially, $\CL_i^\prime$ can
be as small as a singleton, containing only the column representing
not caching the item at all. One iteration of column generation solves
the following LP.

\begin{subequations}
\label{eq:rmp}
\begin{align}
\label{eq:rmpobj}
  \qquad & \hspace{-4mm} \max_{\bm{\chi}} \limits\
  \sum_{i \in \CI} \sum_{\ell \in \CL_i} f_{i\ell} \chi_{i\ell} \qquad &  \\
\label{eq:rmpone}
\qquad &   \text{s.t.}  \ \
\sum_{\ell \in \CL_i} \chi_{i\ell} = 1, \ i \in \CI  & \\
\label{eq:rmpccap}
\qquad &
\hspace{2mm} \sum_{i \in \CI} \sum_{\ell \in \CL_i} b_{it\ell} s_i \chi_{i\ell} \leq C, \ t \in \CT  & \\
\label{eq:rmpbcap}
\qquad &
\hspace{2mm} \sum_{i \in \CI} \sum_{\ell \in \CL_i} u_{it\ell} s_i \chi_{i\ell} \leq L, \ t \in \CT  & \\
\label{eq:rmpchidomain}
\qquad &
\hspace{2mm} 0 \leq \chi_{i\ell} \leq 1, \ i \in \CI, \ell \in \CL_i  &
\end{align}
\end{subequations}

At the optimum of \eqref{eq:rmp}, denote by $\lambda_i^*~(i \in \CI)$,
$\pi_t^*~(t \in \CT)$, and $\mu_t^*~(t \in \CT)$ the corresponding optimal
dual variable values of \eqref{eq:rmpone}, \eqref{eq:rmpccap} and
\eqref{eq:rmpbcap}, respectively.  For any item $i$, the LP reduced
cost of column $\ell$ equals $f_{i\ell}-\lambda_i^* - \sum_{t \in \CT}
b_{it\ell}s_i \pi_t^* - \sum_{t \in \CT} u_{it\ell}s_i \mu_t^*$.  We are interested in knowing if
there exists any column with a positive reduced cost (among those in
$\CL_i \setminus \CL_i^\prime$ for item $i$). If so, adding the column to
\eqref{eq:rmp} shall improve the objective function value.
This can be accomplished by finding the column of maximum
reduced cost for each item. Clearly, the resulting optimization
task decomposes by item.

Recall that any column $\ell$ is characterized by a vector of tuples of binary
values $[(b_{i1\ell}, u_{i1\ell}), (b_{i2\ell}, u_{i2\ell}), \dots,
(b_{iT\ell}, u_{iT\ell})]$. Finding the column of maximum reduced cost
amounts to setting binary values of the tuples, such that the
corresponding reduced cost is maximized.  These values represent the
update and caching decisions of one item, and hence we can use the
variable definitions in Section~\ref{sec:integer}.
The subproblem for a generic item $i$ is formulated below.

\begin{subequations}
\label{eq:sub}
\begin{align}
\label{eq:subobj}
  \qquad & \hspace{-11mm}(\text{SP}_i)~~~~v_i^* = \max_{\bm{x}, \bm{y}} \limits\
  \sum_{t \in \CT} \sum_{a=0}^{t-1} f_i(a) x_{ita} - s_i \sum_{t \in \CT} (\pi_t^* y_{it} + \mu_t^* x_{it0}) &  \\
\label{eq:substate}
\qquad &   \text{s.t.}  \ \
\sum_{a=0}^{t-1} x_{ita} = y_{it}, \ t \in \CT  &\\
\label{eq:subnextslot}
\qquad &
\hspace{2mm} (1-y_{i(t+1)}) + x_{i(t+1)0} + x_{i(t+1)(a+1)} \geq x_{ita},  \ \nonumber \\
& ~~~~~~~~~~t=1,\dots,T-1, a=1,\dots,t-1 &\\
\label{eq:subxdomain}
\qquad &
\hspace{2mm} x_{ita} \in \{0, 1\}, \ t \in \CT, a=0,\dots,t-1 & \\
\label{eq:subydomain}
\qquad &
\hspace{2mm} y_{it} \in \{0, 1\}, \ t \in \CT &
\end{align}
\end{subequations}

There are clear similarities between \eqref{eq:sub} and \eqref{eq:ilp},
as both concern optimizing caching and updating decisions.
However, they differ in several significant aspects.
First, \eqref{eq:sub} deals with a single item.
Second, the cache and backhaul capacities are
not present in \eqref{eq:sub} because these are addressed in
\eqref{eq:reform}. Finally, the objective function \eqref{eq:subobj}
contains the dual variables as the purpose is to maximize the
reduced cost. Note that dual variable $\lambda_i$ is a constant
for item $i$ and hence it is not explicitly included in \eqref{eq:subobj}.

\begin{figure*}
\centering
  \includegraphics[width=0.7\textwidth,height=4cm]{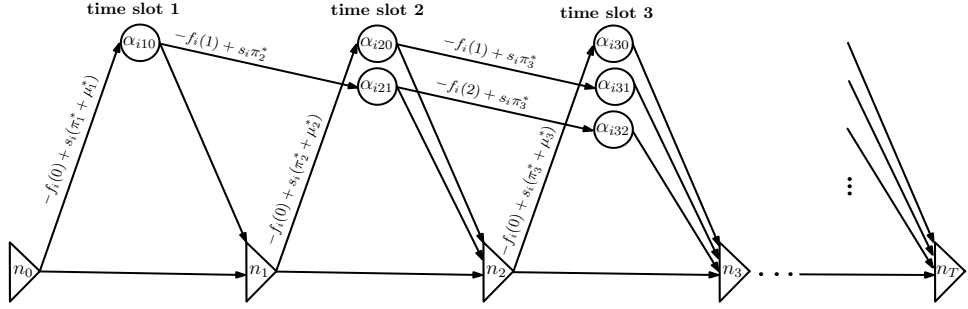}
  \caption{The graph for which finding the shortest path gives the optimum of \eqref{eq:sub}.}
  \label{fig:sp}
\end{figure*}

Subproblem \eqref{eq:sub} is an integer linear problem.  What is less
obvious is that it can be transformed into a shortest path problem
that is polynomial-time solvable. This is illustrated in
Fig.~\ref{fig:sp}, which is a modified single-item version of the
network in Fig.~\ref{fig:flow}. Moreover, the dual variables appear
as part of the arc costs in Fig.~\ref{fig:sp}.

\begin{theorem}
The shortest path from node $n_0$ to $n_T$ gives the optimal solution of \eqref{eq:sub}.
\end{theorem}
\begin{IEEEproof}
  By construction and the proof of Lemma~\ref{theo:fromflow}, a flow
  solution corresponds to a caching solution of the item in question.
  Since the graph is acyclic, a flow solution is a single path.
  Moreover, it is easy to verify that the total path cost equals
  \eqref{eq:subobj} of the corresponding caching solution, and the
  result follows.
\end{IEEEproof}

\begin{remark}
  Suppose a partial decision is taken such that an item shall not be
  cached at a slot. This amounts to simply deleting all the arcs
  entering the nodes representing the age values for that slot.
  Similarly, the partial decision of caching the item in a slot
  corresponds to deleting the arc representing the opposite.  To
  account for a partial decision of updating the item in a slot
  $t$, we find first the shortest path problem from the source to node
  $n_{t-1}$, and then include arc $(n_{t-1}, \alpha_{it0})$ and find
  the shortest path from $\alpha_{it0}$ to the destination node.  In
  conclusion, solving the subproblem can easily accommodate partial
  decisions. This observation is useful for attaining an integer
  solution (Section~\ref{sec:integrality}). $\square$
\end{remark}

\begin{algorithm}[ht!]
{\bf Input}: $\CI$, $\CT$, $s_i, i \in \CI$, $C$, $L$, $f_i(\cdot), i \in \CI$ \\
{\bf Output}: $\bm{\chi}^*$ \\
\vspace{-4mm}
\caption{Column generation}
\label{alg:cg}
\begin{algorithmic}[1]
\STATE Initialize $\CL_i, i \in \CI$
\STATE Stop $\gets$ False
\WHILE {Stop = False}
\STATE Solve \eqref{eq:rmp} to obtain optimum $\bm{\chi}^*$ and dual optimum $(\bm{\lambda}^*, \bm{\pi}^*, \bm{\mu}^*)$
\STATE Stop $\gets$ True
\FOR   {$i=1,\dots,I$}
    \STATE Solve \eqref{eq:sub}
    \IF {$v_i^* - \lambda_i^* > 0$}
      \STATE Stop $\gets$ False
      \STATE Add the column corresponding to $v_i^*$ to $\CL_i^\prime$
    \ENDIF
\ENDFOR
\ENDWHILE
\end{algorithmic}
\end{algorithm}

Column generation for ACOP is summarized in Algorithm~\ref{alg:cg}.
Applying Algorithm~\ref{alg:cg}, the optimal objective function value is the LP
optimum of \eqref{eq:reform} and is therefore an upper bound (UBD) of
the global optimum of ACOP.  This UBD is very useful to gauge
performance of any suboptimal solution, because the deviation from the
global optimum is bounded by that to the UBD.

\subsection{Attaining Solution Integrality}
\label{sec:integrality}

The solution by Algorithm~\ref{alg:cg} may be fractional. A naive
rounding algorithm would pick some fractional $\chi$-variable of some
item $i$ and round it either up to one or down to zero, depending on
the value. This way of rounding is rather aggressive, however, because the
caching and updating decisions over all time slots become fixed, and
there would be no opportunity to make any further decision for
item $i$.

We consider performing rounding more gracefully.  The idea is to make
a rounding decision for one item and one time slot at a time.
Given the optimum LP solution $\bm{\chi}^*$ via column generation, we
define (cf.\ variables used in formulation \eqref{eq:ilp} in
Section~\ref{sec:integer}) $w_{it} = \sum_{\ell \in \CL_i^\prime} b_{it\ell}
\chi^*_{i\ell}$ and $z_{it} = \sum_{\ell \in \CL_i^\prime} u_{it\ell}
\chi^*_{i\ell}$.  These entities can be interpreted as the likelihood
of caching and updating item $i$ in time slot $t$, respectively. Note
that for item $i$, there may be multiple (and factional)
$\chi^*$-variables contributing to the values of $w_{it}$ and
$z_{it}$. The following theorem states that, to attain an integer
solution for \eqref{eq:rmp}, it is sufficient that $w_{it}$ and
$z_{it}$, $i \in \CI, t \in \CT$, become integer.

\begin{theorem}
\label{theo:integer}
$\bm{\chi}^*$ is integer if and only if $\bm{z}$ and $\bm{w}$
are integer.
\end{theorem}
\begin{IEEEproof}
  The necessity is obvious by the definition of $\bm{z}$ and
  $\bm{w}$.  For sufficiency, assume $\bm{z}$ and $\bm{w}$ are
  integer.  Suppose $\bm{\chi}^*$ is fractional, and assume more
  specifically $0 < \chi_{i\ell}^* < 1$. By \eqref{eq:rmpone}, there
  must exist at least another $\chi$-variable for item $i$ that has
  fractional value. Let ${\hat \CL}_i$ denote the column index set of
  the fractional $\chi$-variables of item $i$. Because there are no
  identical columns in $\CL_i^\prime$, there must exist some time slot $t$, for
  which ${\hat \CL}_i$ can be partitioned into two sets, ${\hat
    \CL}_i^0$ and ${\hat \CL}_i^1$, such that $b_{t\ell}=0, \ell \in
  {\hat \CL}_i^0$ and $b_{t\ell}=1, \ell \in {\hat \CL}_i^1$, and none
  of the two sets is empty.  Thus $w_{it} = \sum_{\ell \in \CL_i^\prime}
  b_{t\ell} \chi^*_{i\ell} = \sum_{\ell \in {\hat \CL}_i^1} b_{t\ell}
  \chi^*_{i\ell} < 1 $, contradicting that $w_{it}$ is integer.
  The same argument applies to $\bm{z}$, and the theorem follows.
\end{IEEEproof}

By the above result, rounding can be performed on $\bm{z}$ and
$\bm{w}$, instead of $\bm{\chi}^*$.  This amounts to considering
disjunctions, i.e., a partition of the $\chi$-variables into two
subsets rather than considering a single $\chi$-variable (cf.\ the
proof of Theorem~\ref{theo:integer}), and assembles the use of
disjunctions in branch-and-bound in solving integer
programs~\cite{KaCo09}.  Hence we use the term disjunction-based
rounding (DR) to refer to to the rounding concept.

After performing DR, column generation is applied again.  This is
because additional columns may be needed to reach the LP optimum given
the constraint imposed by rounding. For example, setting $y_{it} = 0$
means to exclude columns $\{\ell \in \CL_i: b_{it\ell}=1\}$ and since
the current $\CL_i^\prime$ is a small subset of $\CL_i$, some new column $\ell \notin
\CL_i^\prime$ with $b_{it\ell}=0$ may improve the objective function. This
yields the repeated column generation algorithm, or RCGA in short.

Note that, after a DR operation, both \eqref{eq:rmp} and the
subproblem of column generation have to comply with the rounding
decision.  Namely, if $y_{it}=0$ by DR, then $\{\ell \in \CL_i^\prime:
b_{it\ell}=1\}$ are removed from \eqref{eq:rmp}, and arcs representing
caching the item $i$ in slot time $t$ in the subproblem graph are
deleted. If $y_{it}=1$, then $\{\ell \in \CL_i^\prime: b_{it\ell}=0\}$ are removed from
\eqref{eq:rmp}, and the subproblem for item $i$ becomes two (smaller)
shortest path problems, then the arc representing not being in the cache
is deleted. Similar updates apply for DR on $\bm{z}$. Thus the subproblem
remains polynomial-time solvable.

The details of DR is presented in Algorithm~\ref{alg_DR}.  In the
algorithm, symbol~$\leftarrow$ is used to indicate assignment of
value. Symbol~$\leftleftarrows$ is used to indicate that an assigned
value of an optimization variable is kept fixed after the assignment.
Lines~$1$-$2$ calculate $\bm{z}$ and $\bm{w}$. For item
$i\in\mathcal{I}$ and time slot $t\in \mathcal{T}$, Line $3$ makes the
decisions of fixing $x_{it0}=1$ and $y_{it}=1$, if $z_{it}=1$, and
Line~$4$ discards the columns that do not comply to the decisions.
Line $5$ performs the opposite decision of fixing $x_{it0}=0$ and
$y_{it}=0$, if $w_{it}=0$, and the non-complying columns are discarded
(Line~$6$).  Lines~7-10 calculate the remaining spare backhaul and
cache capacities, denoted by $L^\prime$ and $C^\prime$, respectively.

Lines~\ref{fracsz}~and~\ref{fracsw} compute the number of fractional
elements of $\bm{z}$ and $\bm{w}$, denoted by $\xi$ and $\eta$,
respectively. If both $\xi$ and $\eta$ equal to zero, the current
solution is integer by Theorem \ref{theo:integer}.  Otherwise, one DR
is applied for $\bm{z}$ if $\xi>0$ (Lines~14-26). If $\bm{z}$ is
integer but $\eta > 0$, DR is applied to a fractional element of
(Lines~28-42).

More specifically, Lines~\ref{minz}-\ref{argminz} find the fractional
element of $\bm{z}$ being closest to zero, and the corresponding item
and time slot.  These entities are denoted by $\munderbar{z}$,
$\munderbar{i}$, and $\munderbar{t}$, respectively.  The corresponding
entities in examining the fractional element of $\bm{z}$ that is
closest to one (Lines~\ref{maxz}-\ref{argmaxz}) are denoted by
$\bar{z}$, $\bar{i}$, and $\bar{t}$, respectively. If
$\munderbar{z}<\bar{z}$, DR fixes $x_{\munderbar{i}\munderbar{t}0}$ to
be zero in Line~\ref{fix_xunderbarto0}. Furthermore, the non-complying
columns are discarded from $\mathcal{L}_{\munderbar{i}}$ in
Line~\ref{fix_yunderbarto0}. If $\munderbar{z}\ge\bar{z}$, the
algorithm checks whether or not there is enough remaining backhaul
capacity to download item $\bar{i}$. If the answer is positive,
$x_{\bar{i}\bar{t}0}$ is fixed to be one in SP$_{\bar{i}}$ by
Line~\ref{fix_xbarto1}, and the non-complying columns are deleted from
$\mathcal{L}_{\bar{i}}$ by Line~\ref{fix_ybarto0}. If the remaining capacity
does not permit to download $\bar{i}$, $x_{\bar{i}\bar{t}0}$
is fixed to be zero by Line~\ref{fix_xbarto0} and the non-complying
columns will be discarded from $\mathcal{L}_{\bar{i}}^\prime$ by
Line~\ref{fix_ybarto01}. DR based on $\bm{y}$ is similar to that for
$\bm{z}$, and the details are presented in
Lines~\ref{start_eta}-\ref{end_eta}.

As the next step, the algorithm updates the remaining spare backhaul
and cache capacity limits (Lines~\ref{compute spare1}-\ref{compute
  spare2}).  Thereafter, the algorithm examines if any content item
has a larger size than what can be admitted by the capacity limit in
any time slot.  For any such item and time slot, the corresponding
variable is fixed to zero, and the non-complying columns are
discarded (Lines~\ref{larger_size1}-\ref{larger_size2}).  Finally, to
ensure feasibility, an auxiliary column is added for each item
(Lines~\ref{addingcol1}-\ref{addingcol2}).  If the item is cached or
updated in a time slot by the decisions made by DR thus far, the
corresponding parameter is set to one. The rest of elements of the
column are zeros.

\begin{algorithm}
\caption{DR Algorithm}
\begin{algorithmic}[1]
\algsetup{linenosize=\tiny}
\small
\STATE Compute $\bm{z}=\{z_{it}, i \in \mathcal{I}, t \in \mathcal{T}\}$, where $z_{it}=\sum_{\ell\in \mathcal{L}_i}u_{it\ell}\chi^*_{il}$
\STATE Compute $\bm{w}=\{w_{it},  i \in \mathcal{I},t \in \mathcal{T}\}$, where $w_{it}=\sum_{\ell\in \mathcal{L}_i}b_{it\ell}\chi^*_{i\ell}$

\STATE  $x_{it0}\leftleftarrows 1$ and $y_{it}\leftleftarrows 1$ in SP$_i$ if $z_{it}=1$, $ i \in \mathcal{I}, t\in \mathcal{T}$ \label{fixxto1}
\STATE  $\chi_{i\ell}\leftleftarrows0$ in RMP if $u_{it\ell}=0$, $i \in \mathcal{I}, t\in \mathcal{T}, \ell \in \mathcal{L}_i^\prime$\label{fixxito01}
\STATE  $x_{it0}\leftleftarrows0$ and $y_{it}\leftleftarrows0$ in SP$_i$ if $w_{it}=0$, $ i \in \mathcal{I}, t\in \mathcal{T}$ \label{fixxto02}
\STATE  $\chi_{i\ell}\leftleftarrows0$ in RMP if $b_{it\ell}=1$, $i \in \mathcal{I}, t\in \mathcal{T}, \ell \in \mathcal{L}_i^\prime$\label{fixxito02}
\STATE $\mathcal{I}^\prime \leftarrow \{i \in \mathcal{I}| x_{it0} \text{ is fixed to one}\}$
\STATE $\mathcal{I}^{\prime\prime} \leftarrow \{i \in \mathcal{I}| y_{it} \text{ is fixed to one}\}$
\STATE $C^\prime\leftarrow C-\sum_{i \in \mathcal{I}^\prime}s_i$
\STATE $L^\prime\leftarrow L-\sum_{i \in \mathcal{I}^{\prime\prime}}s_i$
\STATE $\xi\leftarrow\underset{ i\in \mathcal{I},t\in\mathcal{T}~~~~~~~~~~~~~~~~~~~~~~~~~~~} {\text{cardinality}\{z_{it}| z_{it}>0~\text{and}~z_{it}<1\}}$\label{fracsz}
\STATE $\eta\leftarrow\underset{i\in \mathcal{I},t\in\mathcal{T} ~~~~~~~~~~~~~~~~~~~~~~~~~~~} {\text{cardinality}\{w_{it}| w_{it}>0~\text{and}~w_{it}<1\}}$\label{fracsw}

\IF {$\xi>0$}
\STATE $\munderbar{z}\leftarrow\underset{i\in \mathcal{I},t\in\mathcal{T}~~~~~~~~~~~~~~~~~~~~~~~~~~~} {\min\{z_{it}| z_{it}>0~\text{and}~z_{it}<1\}}$\label{minz}
\STATE $(\munderbar{t},\munderbar{i})\leftarrow\underset{ i\in \mathcal{I},t\in\mathcal{T}~~~~~~~~~~~~~~~~~~~~~~~~~~~~~~~} {\arg\min\{z_{it}| z_{it}>0~\text{and}~z_{it}<1\}}$\label{argminz}

\STATE $\bar{z}\leftarrow\underset{ i\in \mathcal{I},t\in\mathcal{T}~~~~~~~~~~~~~~~~~~~~~~~~~~~~~~~~} {\min\{1-z_{it}| z_{it}>0~\text{and}~z_{it}<1\}}$\label{maxz}
\STATE $(\bar{t},\bar{i})\leftarrow\underset{i\in \mathcal{I},t\in\mathcal{T}~~~~~~~~~~~~~~~~~~~~~~~~~~~~~~~~~~~~} {\arg\min\{1-z_{it}| z_{it}>0~\text{and}~z_{it}<1\}}$\label{argmaxz}
\IF {$(\munderbar{z}<\bar{z})$} \label{nearestcheck_z}
\STATE $x_{\munderbar{t}\munderbar{i}0} \leftleftarrows 0$ in  SP$_{\munderbar{i}}$\label{fix_xunderbarto0}
\STATE  $\chi_{\munderbar{i}\ell}\leftleftarrows0$ if $u_{\munderbar{i}\munderbar{t}\ell}=1$, $\ell \in \mathcal{L}_{\munderbar{i}}^\prime$\label{fix_yunderbarto0}

\ELSIF{$(s_{\bar{i}}\le L^\prime)$}
\STATE  $x_{\bar{i}\bar{t}0}\leftleftarrows1$ in SP$_{\bar{i}}$\label{fix_xbarto1}
\STATE   $\chi_{\bar{i}\ell}\leftleftarrows0$ if $u_{\bar{i}\bar{t}\ell}=0$, $\ell \in \mathcal{L}_{\bar{i}}^\prime$\label{fix_ybarto0}
\ELSE
\STATE  $x_{\bar{i}\bar{t}0}\leftleftarrows0$ in SP$_{\bar{i}}$\label{fix_xbarto0}
\STATE   $\chi_{\bar{i}\ell}\leftleftarrows0$ if $u_{\bar{i}\bar{t}\ell}=1$, $\ell \in \mathcal{L}_{\bar{i}}^\prime$\label{fix_ybarto01}
\ENDIF
\ELSIF {$\eta>0$}
\STATE   $y_{it}\leftleftarrows1$ in SP$_i$ if $w_{it}=1$, $ i \in \mathcal{I},t \in \mathcal{T}$ \label{start_eta}
\STATE   $\chi_{i\ell}\leftleftarrows0$ in RMP if $b_{it\ell}=0$, $i \in \mathcal{I}, t \in \mathcal{T}, \ell \in \mathcal{L}_i^\prime $
\STATE $\munderbar{w}\leftarrow\underset{i\in \mathcal{I},t\in\mathcal{T}~~~~~~~~~~~~~~~~~~~~~~~~~~~} {\min\{w_{it}| w_{it}>0~\text{and}~w_{it}<1\}}$\label{minw}
\STATE $(\munderbar{t},\munderbar{i})\leftarrow\underset{ i\in \mathcal{I},t\in\mathcal{T}~~~~~~~~~~~~~~~~~~~~~~~~~~~~~~~} {\arg\min\{w_{it}| w_{it}>0~\text{and}~w_{it}<1\}}$\label{argminw}

\STATE $\bar{w}\leftarrow\underset{ i\in \mathcal{I},t\in\mathcal{T}~~~~~~~~~~~~~~~~~~~~~~~~~~~~~~~~} {\min\{1-w_{it}| w_{it}>0~\text{and}~w_{it}<1\}}$\label{maxw}
\STATE $(\bar{i},\bar{t})\leftarrow\underset{i\in \mathcal{I},t\in\mathcal{T}~~~~~~~~~~~~~~~~~~~~~~~~~~~~~~~~~~~~} {\arg\min\{1-w_{it}| w_{it}>0~\text{and}~w_{it}<1\}}$\label{argmaxw}
\IF {$(\munderbar{w}<\bar{w})$} \label{nearestcheck_w}
\STATE  $y_{\munderbar{i}\munderbar{t}}\leftleftarrows0$ in SP$_{\munderbar{i}}$
\STATE   $\chi_{\munderbar{i}\ell}\leftleftarrows0$ if $b_{\munderbar{i}\munderbar{t}\ell}=1$, $\ell \in \mathcal{L}_{\munderbar{i}}^\prime$

\ELSIF{$(s_{\bar{i}}\le C^\prime)$}
\STATE  $y_{\bar{i}\bar{t}}\leftleftarrows1$ in SP$_{\bar{i}}$\label{fixxbarto1}
\STATE   $\chi_{\bar{i}\ell}\leftleftarrows0$ if $b_{\bar{i}\bar{t}\ell}=0$, $\ell \in \mathcal{L}_{\bar{i}}^\prime$
\ELSE
\STATE  $y_{\bar{i}\bar{t}}\leftleftarrows0$ in SP$_{\bar{i}}$
\STATE   $y_{\bar{i}\ell}\leftleftarrows0$ if $b_{\bar{i}\bar{t}\ell}=1$, $\ell \in \mathcal{L}_{\bar{i}}^\prime$\label{end_eta}
\ENDIF
\ENDIF

\FOR{$t=1$ ~to~ $T$}
\STATE $\mathcal{I}^\prime \leftarrow \{i \in \mathcal{I}| x_{it0} \text{ is fixed to one}\}$\label{compute spare1}
\STATE $\mathcal{I}^{\prime\prime} \leftarrow \{i \in \mathcal{I}| y_{it} \text{ is fixed to one}\}$
\STATE $C^\prime\leftarrow C-\sum_{i \in \mathcal{I}^\prime}s_i$
\STATE $L^\prime\leftarrow L-\sum_{i \in \mathcal{I}^{\prime\prime}}s_i$\label{compute spare2}
\FOR{$i \in \mathcal{I}\backslash \mathcal{I}^\prime$}\label{larger_size1}
\IF {$s_i> L^\prime$}
\STATE  $x_{it0}\leftleftarrows0$ in SP$_{i}$
\STATE   $\chi_{i\ell}\leftleftarrows0$ in RMP if $u_{it\ell}=1$, $\ell \in \mathcal{L}_{i}^\prime$
\ENDIF
\ENDFOR
\FOR{$i \in \mathcal{I}\backslash \mathcal{I}^{\prime\prime}$}
\IF {$s_i> C^\prime$}
\STATE  $y_{it}\leftleftarrows0$ in SP$_{i}$
\STATE   $\chi_{i\ell}\leftleftarrows0$ in RMP if $b_{it\ell}=1$, $\ell \in \mathcal{L}_{i}^\prime$
\ENDIF
\ENDFOR\label{larger_size2}
\ENDFOR
\FOR{$i=1$ ~to~ $I$} \label{addingcol1}
\STATE Add to $\mathcal{L}_{i}$ column $\ell$,
where $b_{it\ell}=1$ if $x_{it0}$ has been fixed to be one, and $b_{it\ell}=0$ otherwise,
and $u_{it\ell}=1$ if $y_{it}$ has been fixed to be one, and $u_{it\ell}=0$ otherwise, $t \in \mathcal{T}$ \label{addingcol2}
\ENDFOR
\end{algorithmic}
\label{alg_DR}
\end{algorithm}

\subsection{Algorithm Summary}

The framework of RCGA is shown in Algorithm~\ref{alg_CGAandDR1} that
iterates between CGA and DR.  As there are $I$ items and $T$ time
slots, and at least one element of $x$ and $y$ becomes fixed in value
in each iteration, Algorithm~\ref{alg_CGAandDR1} terminates in at most
$I\times T$ iterations.

\begin{algorithm}
\caption{Framework of RCGA}
\label{alg_CGAandDR1}
\begin{algorithmic}[1]
\algsetup{linenosize=\tiny}
\small
\STATE STOP $\leftarrow 0$
\WHILE{(STOP$=0$)}
\STATE Apply Algorithm~2 to obtain $\bm{\chi}^*$ subject to DR decisions made
\IF {($\bm{\chi}^*$ is an integer solution)}
\STATE STOP $\leftarrow 1$
\ELSE
\STATE Apply Algorithm~3
\ENDIF
\ENDWHILE
\end{algorithmic}
\end{algorithm}

\section{Performance Results}

In this section, we present performance evaluation results of RCGA
and the greedy algorithm (GA). We consider ACOP instances of both
small and large sizes. For the former, we compare the utility achieved
by RCGA and GA to the global optimum obtained from solving
ILP~\eqref{eq:ilp}.  By using global optimum as the reference, we
obtain accurate evaluation in terms of the (relative) deviation from
the optimum, referred to as the optimality gap.  For large-size
ACOP instances, it is computationally difficult to obtain global
optimum. Instead, we use the UBD derived from the first iteration of
RCGA as the reference value.  This is a valid comparison because the
deviation with respect to the global optimum does not exceed the deviation from the UBD.
We will see that, numerically, using the UBD remains accurate in
gauging the optimality gap.

We have used ten utility functions from the literature
\cite{SuBiYaKoSh17,8764465,5062058} including linear and non-linear
functions to model the utility of content items with respect to AoI.
The sizes of content items are generated within interval $[1,10]$. We have
set the the cache capacity to 50\% of the total size of
content items, i.e., $C=0.5\sum_{i \in \mathcal{I}}s_i$. The capacity
of backhaul link is set to $L=\rho\sum_{i \in \mathcal{I}}s_i$ where
parameter $\rho$ steers the backhaul capacity in relation to the total
size of content items. We will vary parameters $I$, $T$, and $\rho$ and
study their impact on the overall utility.

\begin{figure}[ht!]
\includegraphics[width=0.40\textwidth]{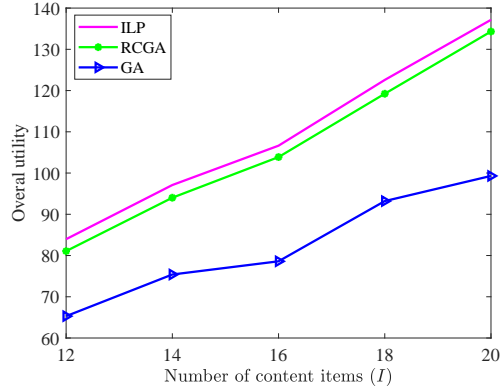}
  \vspace{-2mm}
   \caption{Impact of $I$ on utility when $T=10$, $L=0.3\sum_{i \in \mathcal{I}}s_i$, and $C=0.5\sum_{i \in \mathcal{I}}s_i$.}
  \vspace{-1mm}
  \label{fig:small_I}
\end{figure}

\begin{figure}[ht!]
\includegraphics[width=0.40\textwidth]{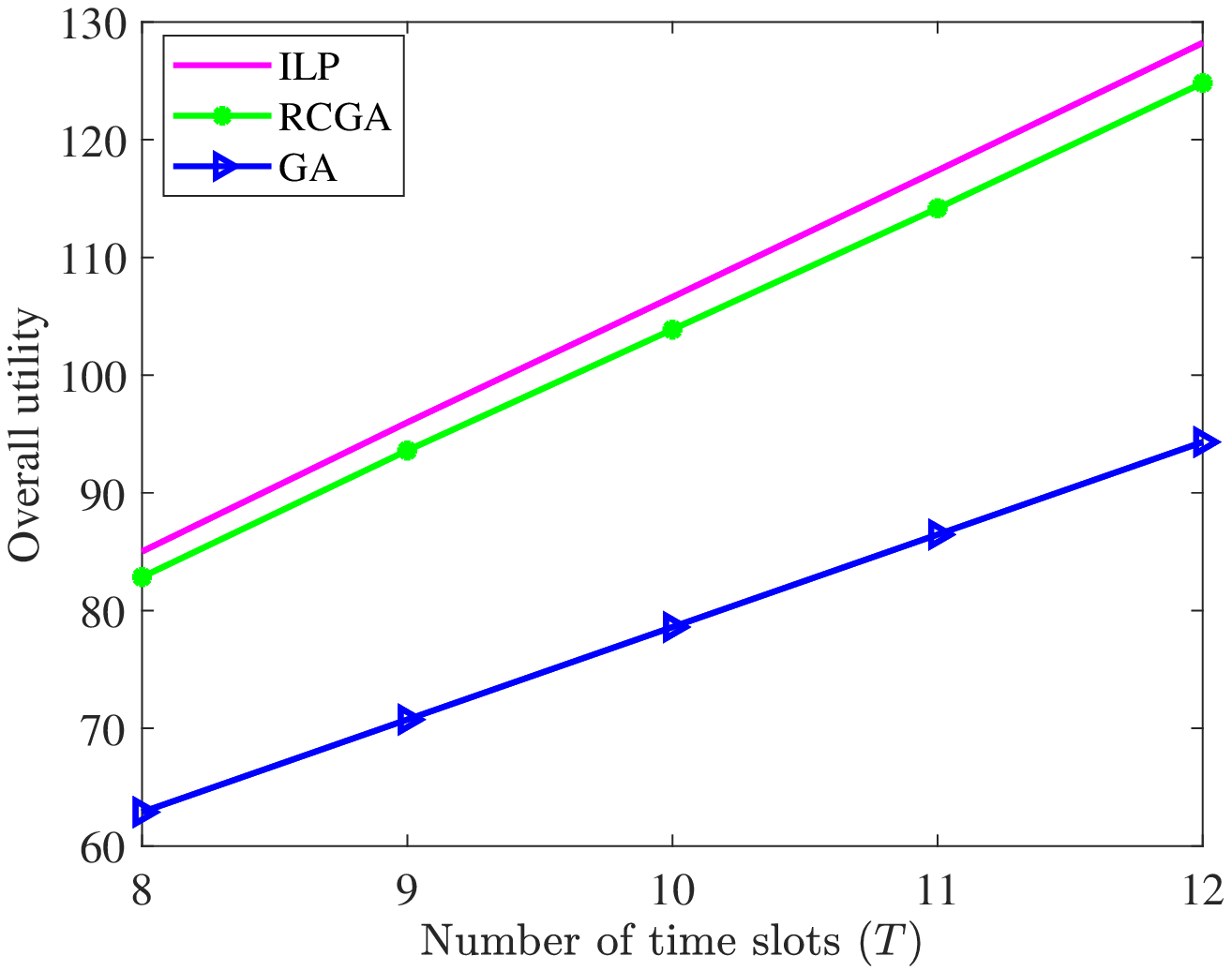}
  \vspace{-2mm}
   \caption{Impact of $T$ on utility when $I=16$, $L=0.3\sum_{i \in \mathcal{I}}s_i$, and $C=0.5\sum_{i \in \mathcal{I}}s_i$.}
  \vspace{-1mm}
  \label{fig:small_T}
\end{figure}

\begin{figure}[ht!]
\includegraphics[width=0.40\textwidth]{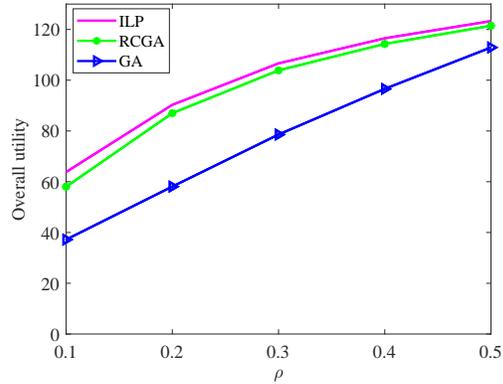}
  \vspace{-2mm}
   \caption{Impact of $\rho$ on utility when $I=16$, $T=10$, and  $C=0.5\sum_{i \in \mathcal{I}}s_i$.}
  \vspace{-1mm}
  \label{fig:small_rho}
\end{figure}

\begin{figure}[ht!]
\includegraphics[width=0.40\textwidth]{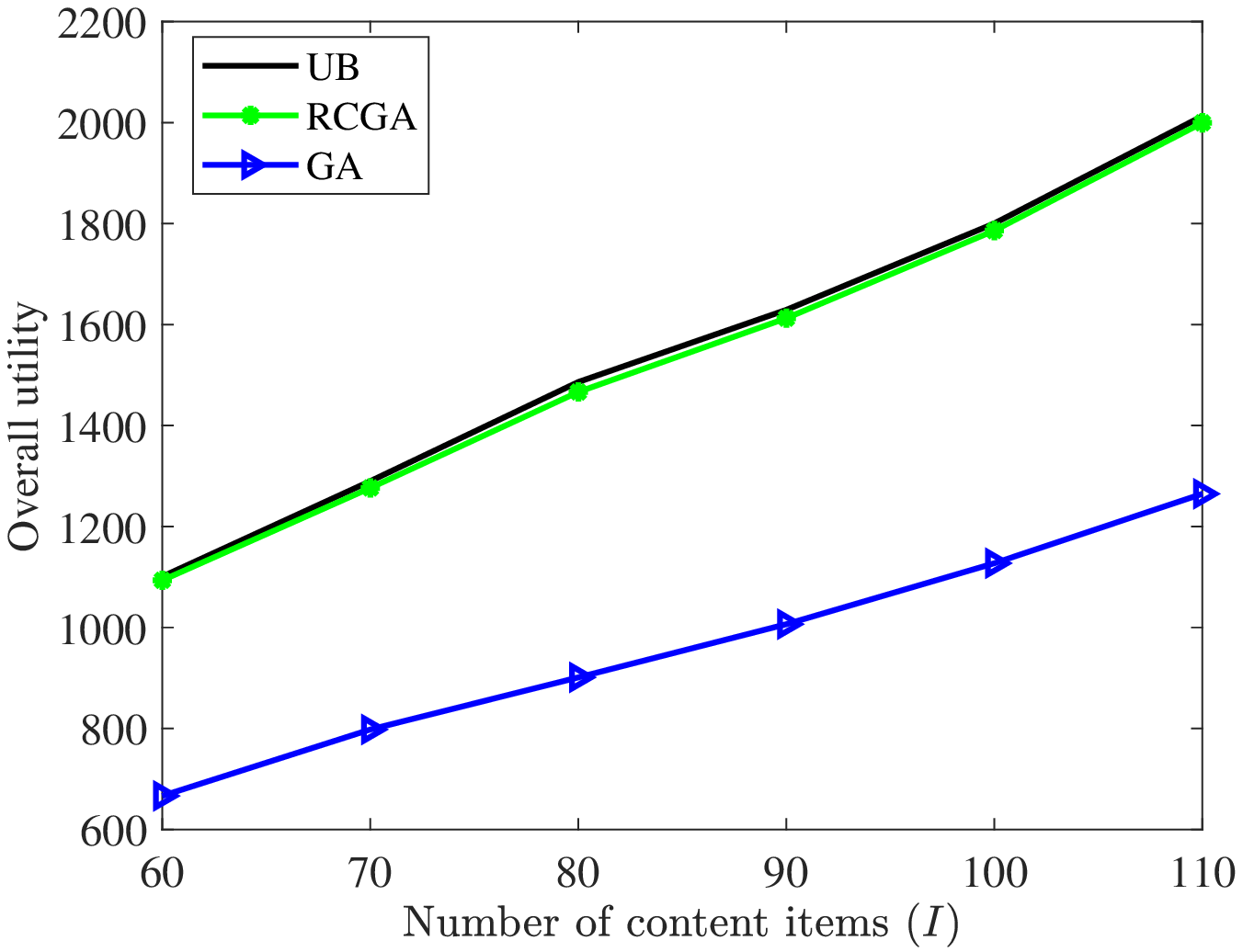}
  \vspace{-2mm}
   \caption{Impact of $I$ on utility when $T=24$, $L=0.3\sum_{i \in \mathcal{I}}s_i$, and $C=0.5\sum_{i \in \mathcal{I}}s_i$.}
  \vspace{-1mm}
  \label{fig:large_I}
\end{figure}

\begin{figure}[ht!]
\includegraphics[width=0.40\textwidth]{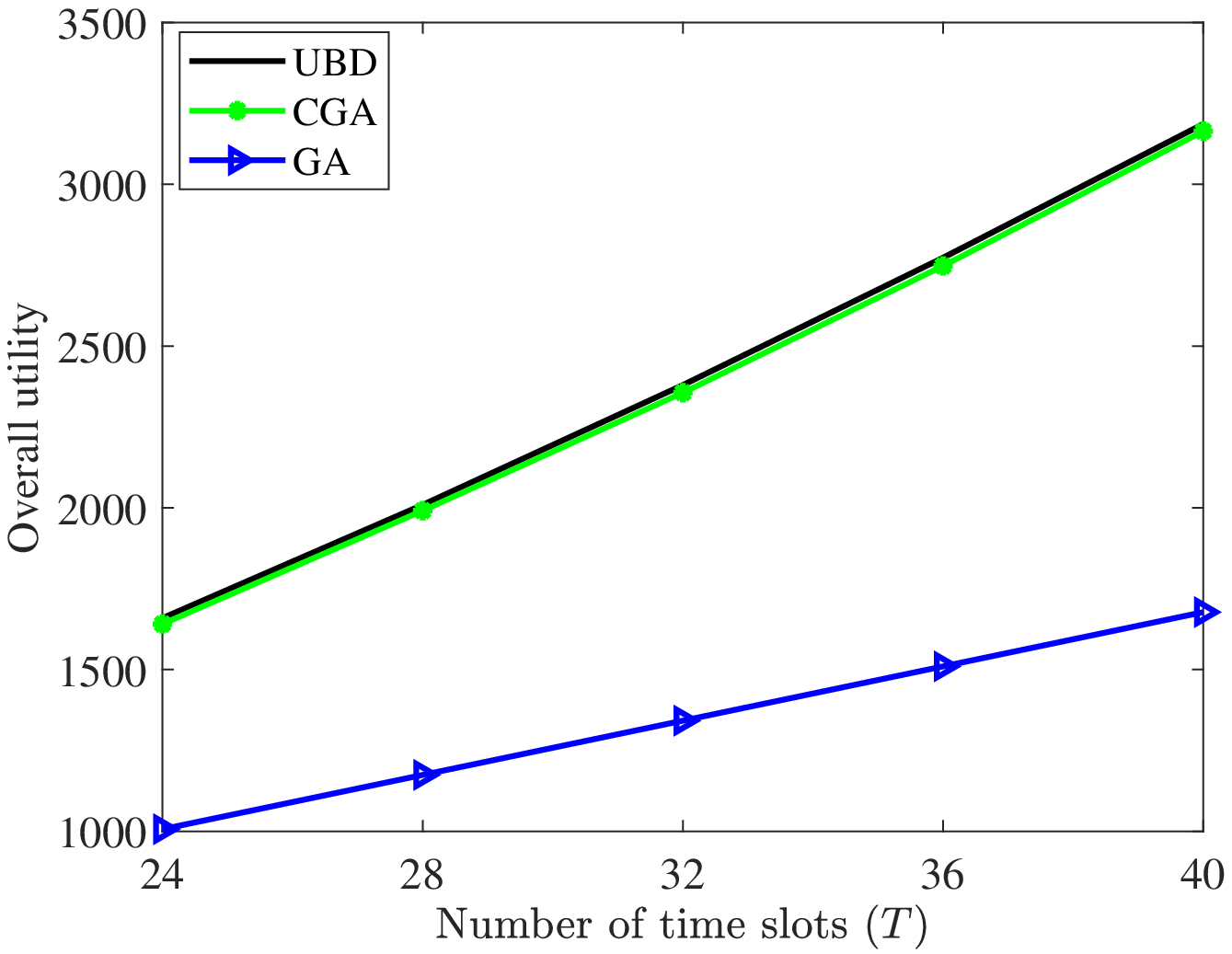}
  \vspace{-2mm}
   \caption{Impact of $T$ on utility when $I=90, L=0.3\sum_{i \in \mathcal{I}}s_i$, and $C=0.5\sum_{i \in \mathcal{I}}s_i$.}
  \vspace{-1mm}
  \label{fig:large_T}
\end{figure}

\begin{figure}[ht!]
\includegraphics[width=0.40\textwidth]{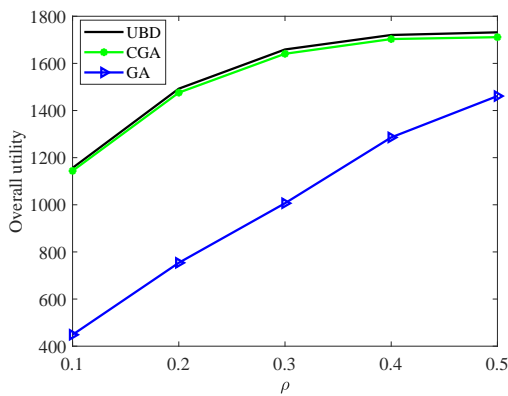}
  \vspace{-2mm}
   \caption{Impact of $\rho$ on utility when $I=90$, $T=24$, and  $C=0.5\sum_{i \in \mathcal{I}}s_i$.}
  \vspace{-1mm}
  \label{fig:large_rho}
\end{figure}

Figs.~\ref{fig:small_I}-\ref{fig:small_rho} and
Figs.~\ref{fig:large_I}-\ref{fig:large_rho} show the performance
results for the small-size and large-size problem instances,
respectively. In Figs.~\ref{fig:small_I}-\ref{fig:small_rho}, the
magenta line represents the global optimum computed using ILP.  In
Figs.~\ref{fig:large_I}-\ref{fig:large_rho} the black line
represents the UBD. In all figures, the green and blue lines represent
the utility achieved by RCGA and GA, respectively.
Overall, RCGA delivers close-optimal solutions (with a few percent of deviation from
optimum). For GA, the deviation from optimality is significantly
larger. Moreover, it can be seen that, the results for small-size
problem instances are consistent with those for larger problem size.
Thus we will mainly discuss the results for small-size problem
instances, even though we will also comment on the difference when the
instance size grows.

Figs.~\ref{fig:small_I} shows the impact of the number of content items
on utility. Apparently, the overall utility increases with the number
of items. This is because when there are more content items, there are
more opportunities to exploit the item-specific utility values in
optimizing the cache.  However GA is less capable of doing so in
comparison to RCGA, as the optimality gap of RCGA is consistently about
3\% only, whereas the optimality gap of GA increases from $22\%$ for
$I=12$ to $28\%$ for $I=20$.

The overall utility will obviously increase for a longer caching time
horizon.  This is seen in Fig.~\ref{fig:small_T}.  RCGA and GA
offer solutions that are approximately 2\% and 26\% from global
optimality for values of $T$ from $8$ to $12$. These optimality gap
values are quite constant over $T$, and the reason is that
extending the time has little structural effect on ACOP.

Fig.~\ref{fig:small_rho} shows the impact of $\rho$ on utility.
Larger $\rho$ means higher backhaul capacity and thus higher utility
as well. There is a saturation effect, however, as when $\rho$
approaches $0.5$ (which corresponds to the cache capacity), the
backhaul capacity hardly constrains the performance.
For the lowest $\rho$-value of $0.1$, the optimality
gaps of RCGA and GA are $9\%$ and $42\%$, respectively, showing that
ACOP becomes noticeably harder if only few content items
can be updated per time slot.


The results in Figs.~\ref{fig:large_I}-\ref{fig:large_rho} follow
the same trend as the first three performance figures.  Interestingly,
for large-size ACOP instances, RCGA delivers better performance, as
the achieved utility by RCGA virtually overlaps with UBD.
We believe this is an effect of the knapsack structure of ACOP.  When
the number of content items is small, even few sub-optimal caching and
updating decisions made by RCGA has noticeable impact on the overall
sub-optimality. This is of less an issue with many content items. GA,
however, has a worse performance for larger-size problem instances.
The observation demonstrates the strength of RCGA over the simple
greedy schedule. Finally, our performance evaluation using the UBD is
accurate, because the global optimum is between the utility by RCGA and
the UBD that are almost overlapping.

\section{Extension to Cyclic Schedule}
\label{sec:cyclic}

Suppose the caching and updating schedule is cyclic. Namely, the
schedule repeats itself for every $T$ time slots. Hence the AoI of an
item in time slot one depends on the scheduling decisions made for the
item in later time slots. In this section, we discuss applying our
results and algorithmic notions to cyclic schedule.

The NP-hardness of cyclic scheduling clearly remains, as
the proof of Theorem~\ref{theo:nphard} applies directly. For uniform
item size, the polynomial-time tractability stated in
Theorem~\ref{theo:uniform} can be generalized to cyclic schedule based
on the notion of flow circulation, which refers to a flow pattern
satisfying flow balance at every node of a graph without source or destination
nodes. We modify the network graph (in Fig.~\ref{fig:flow}) such
that the last node $n_T$ coincides with the first node $n_0$.
Moreover, each node $n_i, i=1,\dots,T-1$ is split into two nodes $n_i$
and $n'_i$ connected by a single arc $(n_i, n'_i)$ of tuple $(0, I,
I)$, i.e., there will be exactly $I$ flow units on this arc for every
time slot.  Finding the optimal cyclic schedule corresponds then to
solving the minimum-cost circulation flow problem for which the optimum
can be computed in polynomial time (e.g., \cite{Ta85}).

Adapting the ILP formulation in Section~\ref{sec:integer} to cyclic
scheduling is easy; this amounts to adding an additional constraint of
type \eqref{eq:ilpnextslot} to connect together time slots $1$ and
$T$. The reformulation \eqref{eq:reform} and our algorithm
RCGA remain applicable. The difference from acyclic schedule lies in the
subproblem. Namely, instead of finding a shortest path with graph
construction shown in Fig.~\ref{fig:sp}, the subproblem consists in
finding minimum cost circulation in a modified graph as discussed
above.

Finally, we remark that the underlying rationale of the greedy
solution in Section~\ref{sec:greedy} does not appear very logical for
a cyclic schedule. The greedy algorithm works slot by
slot, and, for each slot, the algorithm uses the AoI values of the
previous slot as input. With cyclic scheduling, this input is not
available as it depends on the whole scheduling solution.
Nevertheless, the greedy acyclic schedule is still a valid solution to
be used as a cyclic schedule, though the true utility values
need to be evaluated afterward.

\section{Conclusions}
\label{sec:conclusion}

We have considered the optimization problem of time-dynamic caching
where the performance metric is age-centric. Our work has led to the
following key findings. First, the problem complexity originates from
the knapsack structure, whereas for uniform item size it is
polynomial-time solvable.  These results settle the boundary of problem
tractability. Second, column generation offers an effective approach
for problem solving in terms of obtaining clear-optimal solutions.  As
another concluding remark, we believe our results and algorithm admit
extensions to a couple of other related performance functions.  One is
time-specific utility function as the popularity of content items may
change over time. The other is the minimization of average AoI.



\bibliographystyle{IEEEtran}
\bibliography{ref}

\end{document}